\documentclass[twocolumn]{aa} 
\usepackage[]{natbib}
\usepackage{graphicx,color}
\usepackage{txfonts}
\usepackage[colorlinks=true, citecolor=blue]{hyperref}
\usepackage[table]{xcolor}

\newcommand{\orcid}[1]{\href{https://orcid.org/#1}{\includegraphics[width=10pt]{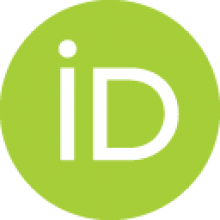}}}

\begin{document} 

\title{CAPOS: The bulge Cluster APOgee Survey III. Spectroscopic Tomography of Tonantzintla~2} 

	\author{
		Jos\'e G. Fern\'andez-Trincado\inst{1}\thanks{To whom correspondence should be addressed; E-mail: jose.fernandez@ucn.cl and/or jfernandezt87@gmail.com}\orcid{0000-0003-3526-5052}, 
		Sandro Villanova\inst{2},
		Doug Geisler\inst{2,3,4},
		Beatriz Barbuy\inst{5}, 
		Dante Minniti\inst{6,7}\orcid{0000-0002-7064-099X},
		Timothy C. Beers\inst{8}\orcid{0000-0003-4573-6233},
		Szabolcs~M{\'e}sz{\'a}ros\inst{9,10,11},
		Baitian Tang\inst{12},
        Roger E. Cohen\inst{13},
		Cristian Moni Bidin\inst{1},
		Elisa R. Garro\inst{6}\orcid{0000-0002-4014-1591}, 
		Ian Baeza\inst{2}\orcid{0000-0002-9881-6336}
		 \and
        Cesar Mu\~noz\inst{3,4}
}
	
	\authorrunning{Jos\'e G. Fern\'andez-Trincado et al.} 
	
\institute{
		Instituto de Astronom\'ia, Universidad Cat\'olica del Norte, Av. Angamos 0610, Antofagasta, Chile
		\and
		Departamento de Astronom\'\i a, Casilla 160-C, Universidad de Concepci\'on, Concepci\'on, Chile
		\and
		Department of Astronomy - Universidad de La Serena - Av. Juan Cisternas, 1200 North, La Serena, Chile
		\and
		Instituto de Investigaci\'on Multidisciplinario en Ciencia y Tecnolog\'ia, Universidad de La Serena. Benavente 980, La Serena, Chile		
		\and
	    Universidade de S\~ao Paulo, IAG, Rua do Mat\~ao 1226, Cidade Universit\'aria, S\~ao Paulo 05508-900, Brazil
		\and
	     Depto. de Cs. F\'isicas, Facultad de Ciencias Exactas, Universidad Andr\'es Bello, Av. Fern\'andez Concha 700, Las Condes, Santiago, Chile
        \and
         Vatican Observatory, V00120 Vatican City State, Italy		
		\and
         Department of Physics and JINA Center for the Evolution of the Elements, University of Notre Dame, Notre Dame, IN 46556, USA
         \and
         ELTE E\"otv\"os Lor\'and University, Gothard Astrophysical Observatory, 9700 Szombathely, Szent Imre H. st. 112, Hungary
        \and
        MTA-ELTE Exoplanet Research Group
        \and 
         MTA-ELTE Lend{\"u}let Milky Way Research Group, Hungary
        \and
        School of Physics and Astronomy, Sun Yat-sen University, Zhuhai 519082, China     
        \and
        Space Telescope Science Institute, 3700 San Martin Dr., Baltimore, MD 21218, USA
    }
	
	\date{Received ...; Accepted ...}
	\titlerunning{Spectroscopic Tomography of Tonantzintla~2}
	
	
	\abstract
	{
		We have performed the first detailed spectral analysis of red giant members of the relatively high-metallicity globular cluster (GC) Tononzintla~2 (Ton~2) using high-resolution near-infrared spectra collected with the Apache Point Observatory Galactic Evolution Experiment II survey (APOGEE-2), obtained as part of the bulge Cluster APOgee Survey. We investigate chemical abundances for a variety of species including the light-, odd-Z, $\alpha$-, Fe-peak, and neutron-capture elements from high S/N spectra of seven giant members. The derived mean cluster metallicity is [Fe/H]$=-0.70\pm0.05$, with no evidence for an intrinsic metallicity spread. Ton~2 exhibits a typical $\alpha$-enrichment that follows the trend for high-metallicity Galactic GCs, similar to that seen in 47~Tucanae and NGC~6380. We find a significant nitrogen spread ($>0.87$ dex), and a large fraction of nitrogen-enriched stars that populate the cluster. Given the relatively high-metallicity of Ton~2, these nitrogen-enriched stars are well above the typical Galactic levels, indicating the prevalence of the multiple-population phenomenon in this cluster which also contains several stars with typical low, first-generation N abundances. We also identify the presence of [Ce/Fe] abundance spread in Ton~2, which is correlated with the nitrogen enhancement, indicating that the \textit{s}-process enrichment in this cluster has been produced likely by relatively low-mass Asymptotic Giant Branch stars. Furthermore, we find a mean radial velocity of the cluster, $-178.6\pm0.86$ km s$^{-1}$ with a small velocity dispersion, 2.99$\pm$0.61 km s$^{-1}$, which is typical of a GC. We also find a prograde bulge-like orbit for Ton~2 that appears to be radial and highly eccentric. Finally, the considerably nitrogen-enhanced population observed in Ton~2, combined with its dynamical properties, makes this object a potential progenitor for the nitrogen-enriched field stars identified so far toward the bulge region at similar metallicity.  
		}
	
	\keywords{stars: abundances -- stars: chemically peculiar -- Galaxy: globular clusters: individual: Ton2 -- techniques: spectroscopic}
	\maketitle

	\section{Introduction}
	\label{section1}
	
	\begin{figure}
		\begin{center}
			\includegraphics[width=90mm]{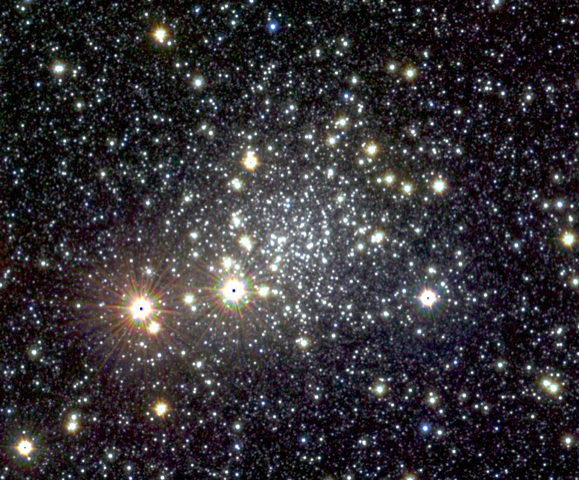}
			\caption{{VVV image of Ton~2}. JHK$_{\rm s}$ colour image centered on Ton~2, with a field of view of 6.5$\times$4.5 arcmin; oriented in Galactic coordinates, with Galactic longitude increasing to the left and Galactic North on top.}
			\label{Figure0}
		\end{center}
	\end{figure}		
	
Galactic Globular Clusters (GCs) are valuable ancient relics of the early epochs of the formation and assembly of the Milky Way (MW). Thus, studies of them both individually as well as an ensemble offers  crucial pieces of information on the chemical evolution and dynamical history of their host galaxies. 
	 
While they were initially assumed to be chemically homogenous, it has been known for decades that they display light-element abundance variations \citep[for reviews, see ][]{Gratton2012, Bastian2018}, which are now termed multiple populations. Almost all the Galactic GCs spectroscopically examined so far have been shown to exhibit star-to-star variations in the light- (C, N), odd-Z (Na, Al), and $\alpha$- (O, Mg) elements \citep{Gratton2004, Carretta2009a, Carretta2009b, Martell2009, Carretta2010, Villanova2010, Villanova2014, Pancino2017, Schiavon2017, Tang2018, Masseron2019, Meszaros2020,Meszaros2021, Frelijj2021, Geisler2021, Romero-Colmenares2021, Tang2021}, as well as variations in the \textit{s}-elements \citep[e.g.,][]{Yong2008, Yong2009, Marino2009, Marino2015}, which seem to be related with the presence of a class of anomalous GCs with peculiar ``chromosome maps" \citep[][the Type II GCs in ]{Milone2017, Marino2019}, leading to the suggestion of a complex chemical-enrichment history. Such variations have been generally hypothesized to be the result of chemical feedback from an earlier population of stars \citep[see, e.g.,][]{Gratton2001, Cohen2002, Meszaros2020}. However, to date this interpretation is still unclear and the nature of any polluters remains uncertain \citep[see, e.g.,][]{Renzini2015, Bastian2018}.

The Apache Point Observatory Galactic Evolution Experiment \citep[APOGEE;][]{Majewski2017}  of the Sloan Digital Sky Survey-IV \citep{Blanton2017} has made major contributions in this field. \citet{Meszaros2015}, \citet{Masseron2019}, \citet{Nataf2019}, \citet{Meszaros2020} and \citet{Meszaros2021} have provided a large homogenous chemical analysis of GCs in the \textit{H}-band, mapping the large population of GC stars with enhanced Al and N abundances. These have enabled the confirmation and detection of the multiple-population phenomenon over a wide range of cluster metallicities. APOGEE's near-IR (NIR) spectral range dramatically reduces the effects of dust obscuration and even allowed, for the first time in most cases, observations of the chemical composition for some of the many GCs in the Galactic bulge, which were previously effectively hidden by the presence of high interstellar extinction and stellar crowding toward this region \citep[see, e.g.,][]{Nataf2019,  6522_2019, UKS1_2020,6723_2021, VVVCL001_2021, M54_2021, Gran2021, Romero-Colmenares2021}. Despite this progress, only a handful of relatively high metallicity GCs have been explored in detail \citep[see, e.g.,][]{Johnson2018}, although those that have provide evidence for the prevalence of the multiple-population phenomenon on the metal-rich end \citep[see, e.g.,][]{Schiavon2017, Tang2017, Meszaros2020}. This is because relatively metal-rich GCs are concentrated in the bulge, but the SDSS-IV survey of the bulge did not primarily target bulge globular clusters (BGCs). Indeed, \citet{Meszaros2020} present the APOGEE sample of 44 clusters, of which only 8 are bona fide BGCs according to \citet{Massari2019}. Of these, they dismiss all but 2 from their analysis as either not having a large enough sample of well-observed members or having too high reddening. 

In order to greatly augment SDSS-IV BGC studies, and help explore the multiple population phenomenon at the high metallicity end, the bulge Cluster APOgee Survey \citep[CAPOS;][]{Geisler2021} was implemented as an CNTAC Contributed program (External program), focussed on APOGEE observations of BGCs. CAPOS observed a total of 17 BGCs.  An overview and initial results for the BGCs observed by CAPOS and available in DR16 (Ahumada et al. 2020) are given in Geisler et al. (2021). A second study (Romero-Colmenares et al. 2021) presents results from CAPOS observations of the recently discovered, intriguing BGC FSR-1758.

Here, we investigate another BGC observed by CAPOS: Tonantzintla~2 (Ton~2 or Pismis~26) in order to explore another BGC with relatively high metallicity, whose elemental abundances have not been previously examined, and originally discovered by \citet{Pismis1959}. Ton~2 is located toward the Galactic bulge, at  $\alpha =$ 17:36:10.56, $\delta = -$38:33:10.8, and close to the Galactic plane ($l=350.79683^{\circ}$, $b= -3.42328^{\circ}$; \citealt{Harris1996}) in a region that is affected by large and spatially variable foreground colour excess, with E(B$-$V)$=1.26$ \citep{Bica1996}, persistent over the entire cluster field \citep{Cohen2018}. The metallicity of this cluster is not well determined, with published values ranging from $-0.73$ to $-0.26$ \citep[see, e.g.,][]{Harris1996, Cote1999, Dias2016b, Vasquez2018}, with no high-resolution spectroscopic study so far.

Ton~2 belongs to the select class of GCs that contain high energy sources, having been clearly detected in X-rays and $\gamma$-rays \citep[e.g.,][]{Hertz1985, Verbunt1995}, and also in the radio region \citep[e.g.,][]{Boyles2011}. This cluster has been controversial from the dynamical point of view, having been classified as a low-energy GC \citep{Masseron2019}, as a thick-disk GC by \citet{Perez-Villegas2020}, and as a possible member of the Koala accretion event  by \citet{Forbes2020}. Here we present the first high-resolution spectroscopic study of Ton~2, based on APOGEE-2 observations. In Section \ref{section2}, we describe the observations. In Section \ref{isochrone}, the global properties of the cluster are presented. In Section \ref{abundances}, the employed atmospheric parameters are described. In Section \ref{elements}, the elemental abundances determined with the \texttt{BACCHUS} code are presented. In Section \ref{mass}, we provide a review of the present mass of Ton~2 based on APOGEE-2 $+$ available kinematic data from the literature.  In Section \ref{model}, the orbit of Ton~2 is examined. Conclusions are presented in Section \ref{conclusions}.
	
	\begin{figure*}
	\begin{center}
		\includegraphics[width=190mm]{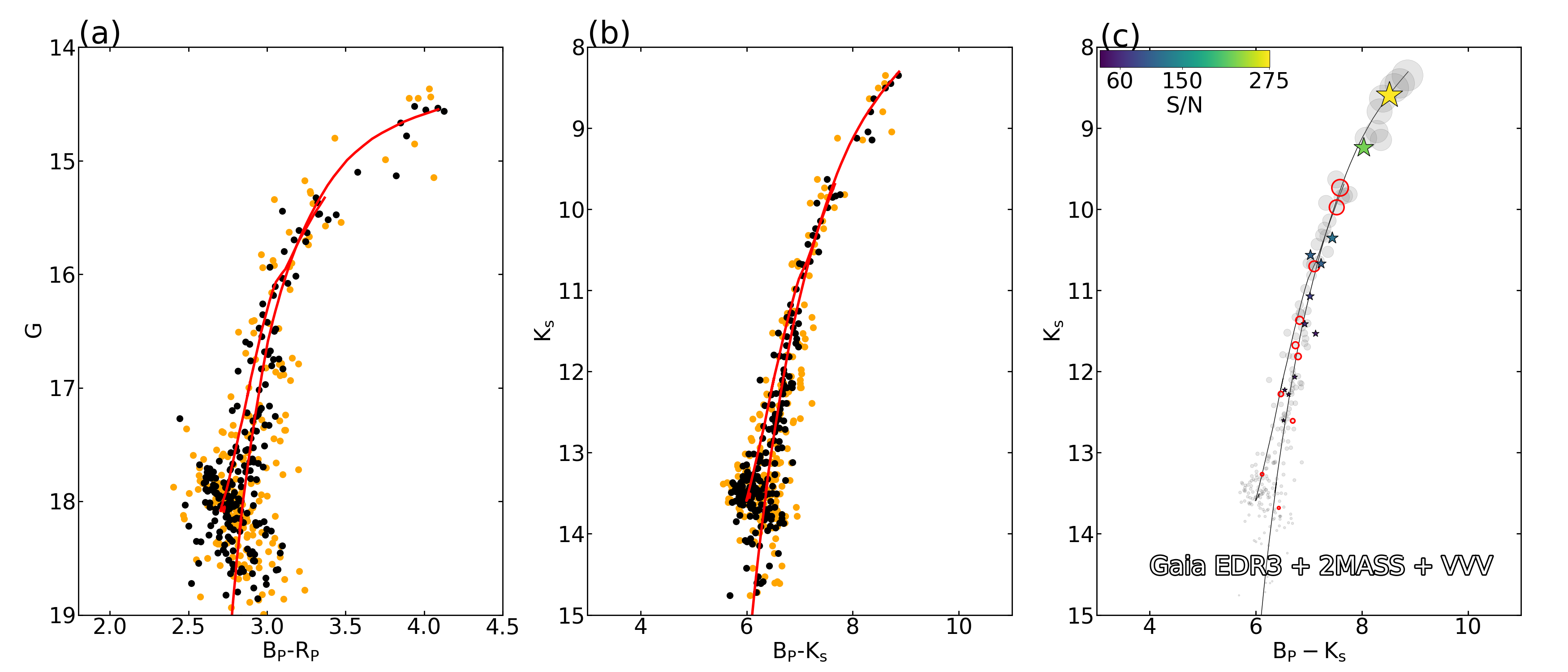}
		\caption{{\bf Differential-reddening corrected (black dots) and uncorrected (orange dots) Optical $+$ NIR CMD}. The best isochrone fit (red line) to Ton~2 stars (black dots) located within 3.5 arcmin from the cluster center for VVV$+$2MASS and \textit{Gaia} EDR3 photometry is shown in panels (a) and (b), respectively, while Ton~2 sources observed with the APOGEE-2 spectrograph are highlighted in the differential-reddening corrected CMD shown in panel (c). The symbol size is according to the brightness  in the K$_{\rm s}$ band, while the  colour-code symbol refers to the S/N of the spectra which is indicated by the inner-top bar. Ton~2 sources with RV information from literature \citep{Baumgardt2019} are highlighted with empty red symbols.}
		\label{Figure1}
	\end{center}
\end{figure*}		
	
\section{Observations}
\label{section2}

We based our analysis on high-resolution ($R\sim22,500$), near-infrared (NIR) spectra taken by the Apache Point Observatory Galactic Evolution Experiment II survey \citep[APOGEE-2;][]{Majewski2017}, one of the internal programs of the Sloan Digital Sky Survey-IV \citep{Blanton2017} developed to provide precise radial velocities (RV $<$1 km s$^{-1}$) and detailed chemical abundances for an unprecedented large sample of giant stars, aiming to unveil the dynamical structure and chemical history of the entire MW galaxy. 

APOGEE-2 observations were carried out through two twin spectrographs \citep{Wilson2019} from the Northern Hemisphere on the 2.5m telescope at Apache Point Observatory \citep[APO, APOGEE-2N;][]{Gunn2006} and the Southern Hemisphere on the Ir\'en\'ee du Pont 2.5m telescope \citep[][]{Bowen1973} at Las Campanas Observatory (LCO, APOGEE-2S). Each instrument records most of the \textit{H}-band (1.51$\mu$m -- 1.69$\mu$m) on three detectors, with coverage gaps between $\sim$1.58--1.59$\mu$m and $\sim$1.64--1.65$\mu$m, and with each fiber subtending a $\sim$2'' diameter on-sky field of view in the northern instrument and 1.3'' in the southern.

DR~17 will be the final release of APOGEE-2 data from SDSS-III/SDSS-IV. It will include all APOGEE-2 data, including data taken at APO through November 2020 and at LCO through January 2021. The dual APOGEE-2 instruments have observed more than $ 650,000$ stars throughout the MW, targeting these objects with selections detailed in \citet{Zasowski2017}, \citet{Beaton2021}, and \citet{Santana2021}. Spectra were reduced as described in \citet{Nidever2015}, and analyzed using the APOGEE Stellar Parameters and Chemical Abundance Pipeline \citep[ASPCAP;][]{Garcia2016}, and the libraries of synthetic spectra described in \citet{Zamora2015}. The customised \textit{H}-band line lists are fully described in \citet{Shetrone2015}, \citet{Hasselquist2016}--neodymium lines (Nd II), \citet{Cunha2017}-- cerium lines (Ce II), and \citet{Smith2021}. 
	
\section{Ton~2}
\label{Ton2}

The GC Ton~2 was observed as part of the bulge Cluster APOgee Survey \citep[][]{Geisler2021}. Figure \ref{Figure0} shows the VISTA Variables in the Via Lactea \citep[VVV;][]{Minniti2010, Saito2012} JHK$_{\rm s}$ colour image, centered on Ton~2 with a field of view of 6.5$\times$4.5 arcmin, revealing the evident density of stars associated with the cluster.

The APOGEE-2S plug-plate containing Ton~2 was centered on ($l$,$b$) $\sim$ (350$^{\circ}$, $-3.0^{\circ}$), the same plug-plate that contains FSR~1758 \citep[see, e.g.,][]{Romero-Colmenares2021}. In this plug-plate 12 of 264 science fibers were positioned for potential Ton~2 members. Targets were selected on the basis of the VVV/VIRAC \citep{Smith2018} $+$ \textit{Gaia} DR2 \citep{Brown2018} $+$ 2MASS \citep{Skrutskie2006} catalogs. Our targets are positioned from near the tip of the red giant branch (RGB), as shown in the differential reddening-corrected colour magnitude diagram (CMD) to several magnitudes fainter. All observed Ton~2 stars had 2MASS \textit{K$_{\rm s}$}-band  brighter than 13. This was required in order to achieve a minimum signal-to-noise, S/N $\gtrsim$60 pixel$^{-1}$, in one plug-plate visit ($\sim$ 1 hour). Although more visits were originally planned, in the end, given weather, time allocation, and airmass constraints, only one visit was indeed obtained. Seven out of the twelve observed stars reached S/N$>60$ pixel$^{-1}$, while the remaining spectra have lower S/N, ranging from 31  to 52 pixel$^{-1}$ (Table \ref{Table1}). In the following, we use all stars to provide reliable and precise ($< 1$ km s$^{-1}$) radial velocities for cluster membership confirmation, but limit ourselves to the seven higher S/N stars for the abundance analysis; these stars  are highlighted in Figure \ref{Figure1}.

	\begin{figure*}
	\begin{center}
		\includegraphics[width=190mm]{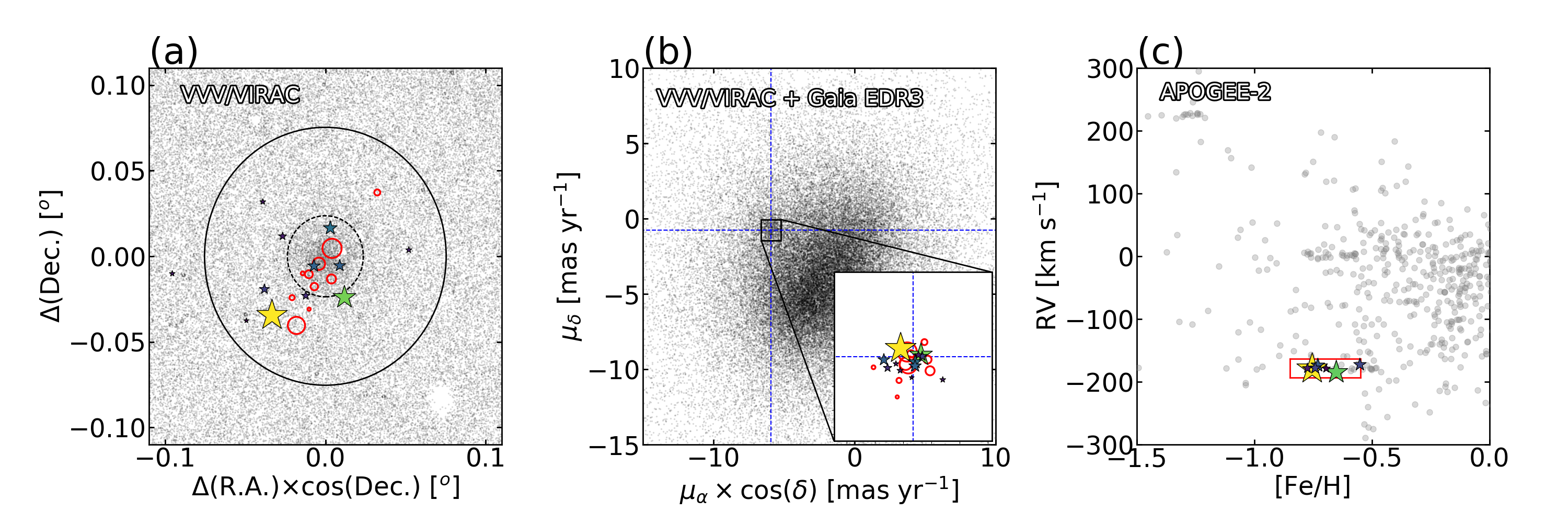}
		\caption{{\bf Global properties of Ton~2.} Panel (a): Spatial position of sources centered on Ton~2, with stars observed by APOGEE-2 highlighted with the same colour-code as Figure \ref{Figure1}, and stars with RV information (red empty circles) from the literature \citep{Baumgardt2019}. The black dashed circle highlights the projected half-light radius of Ton~2, $r_{\rm h,l}=2.89$ pc, while the large black circle marks the two times (2$\times r_{h,m}$) projected half-mass radius ($r_{\rm h,m} = 4.6$ pc) from \citet[][]{Baumgardt2019}. Panel (b): The proper-motion distribution of sources toward the Ton~2 field, with blue dotted lines highlighting the nominal \textit{Gaia} EDR3  proper motions of the cluster \citep[e.g.,][]{Vasiliev2021}. Panel (c): Radial velocity versus metallicity of our members compared to APOGEE-2S field sources toward Ton~2. The [Fe/H] of our targets have been determined with the \texttt{BACCHUS} code (see text), while the [Fe/H] of field stars are from the \texttt{ASPCAP} pipeline. The red box limited by $\pm$0.15 dex and $\pm$15 km s$^{-1}$ and centered on [Fe/H]$= -0.70$ and RV $= -178.61$ km s$^{-1}$ encloses our potential cluster members.}
		\label{Figure2}
	\end{center}
\end{figure*}

\section{Global properties of Ton~2}
\label{isochrone}

The colour-magnitude diagram (CMD) presented in Figure \ref{Figure1} was differential-reddening corrected using giant stars, and by following the same methology as employed in \citet{Romero-Colmenares2021}. For this purpose, we selected all RGB stars within a radius of 3.5 arcmin from the cluster center and that have proper motions compatible with that of Ton~2. First, we draw a ridge line along the RGB, and for each of the selected RGB stars we calculated its distance from this line along the reddening vector. The vertical projection of this distance gives the differential interstellar absorption at the position of the star, while the horizontal projection gives the differential Optical$+$NIR reddening at the position of the star. After this first step, for each star of the field we selected the three nearest RGB stars, calculated the mean interstellar reddening and absorption, and finally subtracted these mean values from its Optical$+$NIR colours and magnitudes. We underline the fact that the number of reference stars used for the reddening correction is a compromise between having a correction affected as little as possible by photometric random error and the highest possible spatial resolution. Figure \ref{Figure1} shows the result of this correction, with black and orange dots highlighting the differential-reddening corrected and uncorrected CMD, respectively, in the \textit{Gaia} EDR3 bands (panel a) and \textit{Gaia} EDR3 $+$ 2MASS bands (panel b). 

In order to estimate the distance of the cluster we performed an isochrone fitting of the RGB using the \texttt{PARSEC} database\footnote{http://stev.oapd.inaf.it/cgi-bin/cmd} \citep{Bre12}. We had to consider that Bulge clusters are affected by high reddening, generally exceeding E(B-V)$>$0.50. The reddening correction highly depends on the spectral energy distribution (SED) of the star, i.e. on its temperature. For this reason, the extinction correction was applied point-by-point to the isochrone using the extinction law of \citet{Car89}. Without considering the SED-reddening dependence we could not have simultaneously fit the RC and the upper-RGB of the cluster (see Figure \ref{Figure1}). The free parameters for this fitting are the true distance modulus, (m-M)$_0$ (or the equivalent distance in pc), the interstellar absorption in the V band, A$_V$, and the reddening-law coefficient, R$_V$. These three parameters were estimated simultaneously using the B$_{\rm p}-$K vs. K and B$_{\rm p}-$R$_{\rm p}$ vs. G CMDs shown in Figure \ref{Figure1}, assuming an age of 12 Gyrs, and a global metallicity that considers the $\alpha$-enhancement of the cluster according to the relation by \citet{Sal93}. [Fe/H] and $\alpha$-enhancement were obtained from the \texttt{BACCHUS} measurements (see Section \ref{elements}). 

We underline the fact that the simultaneous use of visual/blue and infrared filters allows the determination of the extinction-law coefficient R$_V$, which is usually assumed to be 3.1 but that can vary significantly from the canonical value, especially in the direction of the Galactic Bulge \citep{Nataf2016}, where it can easily go down to 2.5. Figure \ref{Figure1} shows that we achieved a very good fit for a distance d$_{\odot}=7.76$ kpc \citep[in reasonable agreement with recent estimations, 6.99 kpc;][]{Baumgardt2021}, an interstellar absorption A$_{\rm V}=4.29$, and an extinction-law coefficient R$_{\rm V}=2.75$. In particular, we confirm that the R$_V$ coefficient in the Bulge direction is lower than the canonical value assumed for other directions in the Galaxy. Assuming an R$_V$ value equal to 3.1 would yield an isochone too blue for the B$_{\rm p}-$R$_{\rm p}$ vs. G CMD. Finally, the interstellar absorption and the extintion-law coefficient we found can be translated to E(B$-$V)$=1.56$ for Ton~2, substantially higher than the foreground interstellar reddening determined by \citet{Bica1996}, E(B$-$V)$=1.26$, and \citet[][]{Harris1996}--Edition 2010, E(B$-$V)$=1.24$.

	\begin{figure*}
	\begin{center}
		\includegraphics[width=180mm]{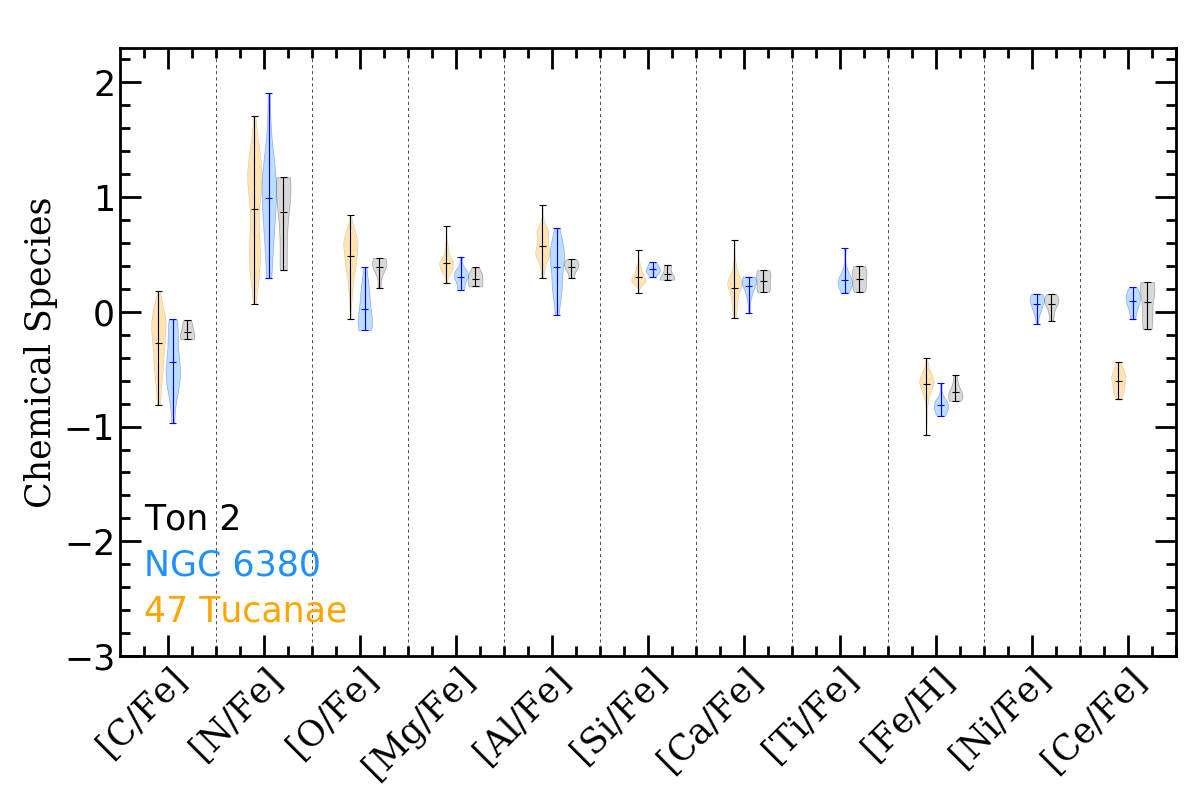}
		\caption{{\bf Elemental abundances of Ton~2}. [X/Fe] and [Fe/H] abundance density estimation (violin representation) of Ton~2 (black), compared to elemental abundances of 47~Tucanae (orange) from \citet{Meszaros2020}, and NGC~6380 stars (blue) from \citet{Fernandez-Trincado2021d}. Each violin representation indicates with vertical lines the mean and limits of the distribution. The abundance ratios shown here have been computed by adopting photometric atmospheric parameters.}
		\label{Figure4}
	\end{center}
\end{figure*}

\section{Kinematic and astrometric properties of Ton~2}

Figure \ref{Figure2} summarizes the main physical properties of our sample. The Ton~2 sources are 
well-positioned within the cluster tidal radius, and most of the sources with high S/N spectra collected with the APOGEE-2 spectrograph are located in the innermost regions of the cluster, e.g., inside $2\times r_{\rm h,m}$ (see Figure \ref{Figure2}(a)). Figure \ref{Figure2}(b) shows the proper-motion distributions of our sources, which are located inside a radius of $\sim$0.7 mas yr$^{-1}$ from the nominal mean proper motions of Ton~2, $\mu_{\alpha}\cos(\delta)= -5.913\pm0.031$ mas yr$^{-1}$ and $\mu_{\delta}=-0.758\pm0.028$ \citep{Vasiliev2021}, which is clearly offset from the main field locus, greatly facilitating cluster membership discrimination.

Figure \ref{Figure2}(c) shows the \texttt{BACCHUS} [Fe/H] abundance ratios versus the radial velocity of our seven potential cluster members compared to field stars with \texttt{ASPCAP}/APOGEE-2 [Fe/H] determinations. The field stars with [Fe/H] determined from \texttt{ASPCAP}/APOGEE-2 have been shifted by $+$ 0.11 dex in order to minimize the systematic differences between APOGEE-2 results and \texttt{BACCHUS} \citep[we refer the reader to][for further details]{Fernnadez-Trincado2020_Aluminum}. The same figure indicates that our targets have very similar velocities, which are extreme compared to the field-star distribution, again supporting cluster membership for all of our targets. 
 
 We find a mean RV from 12 APOGEE-2 stars of $-$178.61$\pm$ 0.86 km s$^{-1}$, which is in reasonable agreement with the value listed in Baumgardt's web service\footnote{\url{https://people.smp.uq.edu.au/HolgerBaumgardt/globular/}}, RV$=-184.72 \pm 1.74$  km s$^{-1}$ \citep{Baumgardt2019}. The red open circles in Figures \ref{Figure2}(a)--(b) refer to 13 stars with RV information compiled  from literature by \citet{Baumgardt2019}, and classified in \citet{Vasiliev2021} as high probability ($>95$ \%) members of  Ton~2.
 
The red box highlighted in Figure \ref{Figure2}(c) encloses the seven potential cluster members with high S/N within $\pm0.15$ dex and $\pm15$ km s$^{-1}$ from the mean [Fe/H]$=-0.70$ and RV$= -178.61$ km s$^{-1}$ of Ton~2, as determined in this work. Other sources that fall inside this box are foreground/background stars with other properties that are not compatible with the cluster. 

Table \ref{Table1} lists the photometric, kinematic, and astrometric properties for likely members of Ton~2. 

\begin{table*}
	\begin{small}
		\begin{center}
			\setlength{\tabcolsep}{0.5mm}  
			\caption{Main physical parameters of Ton~2.}
			\begin{tabular}{cccccccccccc}
				\hline
				\hline
				Gaia-EDR3-Ids & APOGEE-Ids &  $\alpha$      &     $\delta$    &   S/N   & G$_{0}$ & BP$_{\rm 0}$ & RP$_{\rm 0}$&  Ks$_{\rm 0}$  &  RV$\pm \Delta$    & $\mu_{\alpha}\cos(\delta) \pm \Delta$    &  $\mu_{\Delta}$ \\
				&  &  hh:mm:ss      &     dd:mm:ss  &  pixel$^{-1}$   & &    && &  km s$^{-1}$    & mas yr$^{-1}$   &  mas yr$^{-1}$ \\				
				\hline
				\hline
				{\bf S/N $>60$}&     &     &    & &    &   &&  &   & &  \\
				\hline
				\hline
				5961843553827536128 &    2M17360034$-$3835151 &    17:36:00.35 &    $-$38:35:15.2 &    275 &     14.48  &  17.11   &  13.11   &   8.59 &  $-$179.48$\pm$0.02 &     $-$6.02$\pm$0.06 &     $-$0.69$\pm$0.04  \\
				5961843283288451584 &    2M17361421$-$3834371 &    17:36:14.22 &    $-$38:34:37.1 &    224 &     15.05  &  17.27   &  13.65   &   9.24 &  $-$184.86$\pm$0.02 &     $-$5.84$\pm$0.04 &     $-$0.74$\pm$0.03  \\
				5961844142282033664 &    2M17361150$-$3832114 &    17:36:11.50 &    $-$38:32:11.4 &    123 &     15.94  &  17.79   &  14.61   &  10.35 &  $-$173.72$\pm$0.03 &     $-$5.89$\pm$0.05 &     $-$0.80$\pm$0.04  \\
				5961844004803160064 &    2M17360837$-$3833312 &    17:36:08.38 &    $-$38:33:31.2 &    109 &     15.83  &  17.59   &  14.54   &  10.56 &  $-$172.66$\pm$0.03 &     $-$5.89$\pm$0.06 &     $-$0.83$\pm$0.04  \\
				5961843347670794496 &    2M17361331$-$3833304 &    17:36:13.31 &    $-$38:33:30.5 &    105 &     16.04  &  17.90   &  14.75   &  10.67 &  $-$177.99$\pm$0.03 &     $-$6.17$\pm$0.06 &     $-$0.78$\pm$0.04  \\
				5961843656906772736 &    2M17355890$-$3834199 &    17:35:58.90 &    $-$38:34:19.9 &    77  &     16.27  &  18.09   &  15.01   &  11.07 &  $-$178.42$\pm$0.03 &     $-$5.83$\pm$0.06 &     $-$0.75$\pm$0.04  \\
				5961843970483273600 &    2M17360681$-$3834336 &    17:36:06.81 &    $-$38:34:33.7 &    63  &     16.63  &  18.33   &  15.35   &  11.41 &  $-$178.87$\pm$0.04 &     $-$6.13$\pm$0.08 &     $-$0.85$\pm$0.06  \\
				\hline
				\hline
				{\bf S/N $<60$}&     &     &    & &    &    &  && &  & \\
				\hline
				\hline
				5961844618994711296 &    2M17362652$-$3832575 &    17:36:26.53 &    $-$38:32:57.5 &    52  &     17.29  &  18.80   &  15.88   &  12.06 &  $-$180.52$\pm$0.04 &     $-$6.02$\pm$0.12 &     $-$0.87$\pm$0.10  \\
				5961844275382113536 &    AP17360238$-$3832287 &    17:36:02.38 &    $-$38:32:28.7 &    48  &     17.04  &  18.66   &  15.66   &  11.53 &  $-$177.71$\pm$0.06 &     $-$5.86$\pm$0.10 &     $-$0.74$\pm$0.08  \\
				5961847303378573056 &    2M17355860$-$3831162 &    17:35:58.60 &    $-$38:31:16.2 &    46  &     17.05  &  18.77   &  15.86   &  12.22 &  $-$179.83$\pm$0.04 &     $-$5.64$\pm$0.14 &     $-$0.94$\pm$0.10  \\
				5961846783642974080 &    2M17354126$-$3833471 &    17:35:41.26 &    $-$38:33:47.2 &    45  &     17.22  &  18.90   &  15.96   &  12.28 &  $-$179.25$\pm$0.04 &     $-$6.06$\pm$0.11 &     $-$0.81$\pm$0.08  \\
				5961843519467793024 &    2M17355543$-$3835260 &    17:35:55.44 &    $-$38:35:26.1 &    31  &     17.48  &  19.12   &  16.24   &  12.60 &  $-$179.97$\pm$0.08 &     $-$5.92$\pm$0.13 &     $-$0.93$\pm$0.10  \\
				\hline
				\hline
				\citet{Baumgardt2019}&     &     &    & &    && &    &   & &  \\
				\hline
				\hline
				5961842836611880064 &    ...                  &    17:36:05.02 &    $-$38:35:36.2 &    ... &     15.47  &  17.50   &  14.17   &   9.97 &  $-$186.93$\pm$2.60 &     $-$5.95$\pm$0.05 &     $-$0.82$\pm$0.03  \\
				5961843210232686720 &    ...                  &    17:36:07.45 &    $-$38:35:02.0 &    ... &     18.42  &  20.11   &  17.24   &  13.68 &  $-$176.80$\pm$4.41 &     $-$6.05$\pm$0.21 &     $-$1.09$\pm$0.16  \\
				5961843352008650880 &    ...                  &    17:36:11.73 &    $-$38:33:58.5 &    ... &     16.54  &  18.19   &  15.23   &  11.36 &  $-$174.74$\pm$4.28 &     $-$5.76$\pm$0.07 &     $-$0.87$\pm$0.05  \\
				5961843966146050176 &    ...                  &    17:36:04.19 &    $-$38:34:37.8 &    ... &     17.50  &  19.30   &  16.32   &  12.60 &  $-$185.79$\pm$2.69 &     $-$6.03$\pm$0.15 &     $-$0.95$\pm$0.11  \\
				5961844000504170112 &    ...                  &    17:36:07.40 &    $-$38:33:48.7 &    ... &     16.68  &  18.42   &  15.44   &  11.67 &  $-$185.53$\pm$2.29 &     $-$5.78$\pm$0.09 &     $-$0.78$\pm$0.06  \\
				5961844004803196544 &    ...                  &    17:36:06.22 &    $-$38:33:46.8 &    ... &     18.07  &  19.38   &  16.55   &  13.26 &  $-$191.68$\pm$2.11 &     $-$6.26$\pm$0.21 &     $-$0.84$\pm$0.15  \\
				5961844004843014144 &    ...                  &    17:36:08.45 &    $-$38:34:14.6 &    ... &     16.80  &  18.61   &  15.58   &  11.81 &  $-$183.80$\pm$2.74 &     $-$5.93$\pm$0.09 &     $-$0.73$\pm$0.06  \\
				5961844004843042816 &    ...                  &    17:36:09.33 &    $-$38:33:26.5 &    ... &     16.03  &  17.80   &  14.70   &  10.70 &  $-$182.99$\pm$1.89 &     $-$5.97$\pm$0.06 &     $-$0.81$\pm$0.04  \\
				5961844103583415424 &    ...                  &    17:36:11.84 &    $-$38:32:54.1 &    ... &     15.37  &  17.32   &  14.00   &   9.73 &  $-$187.81$\pm$0.34 &     $-$5.96$\pm$0.05 &     $-$0.72$\pm$0.03  \\
				5961845654110501760 &    ...                  &    17:36:20.49 &    $-$38:30:56.6 &    ... &     17.15  &  18.75   &  15.94   &  12.27 &  $-$174.72$\pm$3.36 &     $-$5.81$\pm$0.10 &     $-$0.63$\pm$0.07  \\
				\hline
				\hline
			\end{tabular}  \label{Table1}
		\end{center}
			\raggedright{{\bf Note:} Differential reddening-corrected photometric, kinematic, and astrometric properties for members of Ton~2 are listed.}
	\end{small}
\end{table*}   

\section{Atmospheric parameters}
\label{abundances}

We made use of the Brussels Automatic Stellar Parameter (\texttt{BACCHUS}) code \citep{Masseron2016} to derive the metallicity, broadening parameters, and [X/Fe] abundance ratios for Ton~2 stars by adopting the same technique as described in \citet{6522_2019, UKS1_2020, 6723_2021, M54_2021, VVVCL001_2021} and \citet{Romero-Colmenares2021}. Table \ref{Table2} lists the elemental abundances for the targets analyzed in this work. 

We also applied a simple approach of fixing T$_{\rm eff}$ and $\log$ \textit{g} to values determined independently of spectroscopy, in order to minimize a number of caveats present in \texttt{ASPCAP}/APOGEE-2 abundances for GCs \citep[see, e.g.,][]{Masseron2019, Meszaros2020, Meszaros2021, 6522_2019, UKS1_2020,LMC_SMC_2020, 6723_2021, M54_2021, VVVCL001_2021, Romero-Colmenares2021}. 

In order to obtain T$_{\rm eff}$ and $\log$ \textit{g} from photometry, we first derived the differential-reddening corrected CMD of Figure \ref{Figure1}. We then horizontally projected the position of each observed star until it intersected the isochrone and assumed $T_{\rm eff}$ and $\log$ \textit{g} to be the temperature and gravity of the point of the isochrones that have the same G and/or K$_{\rm s}$ magnitude as the star. We underline the fact that, for highly reddened objects like Ton~2, the interstellar absorption correction depends on the spectral energy distribution of  the star, i.e., on its temperature. For this reason, we applied a temperature-dependent absortion correction to the isochrone. Without this, it is not possible to obtain a proper fit of the RGB, especially of the upper and cooler part. The adopted atmospheric parameters are listed in Table \ref{Table2}.

	\begin{figure*}
	\begin{center}
		\includegraphics[width=190mm]{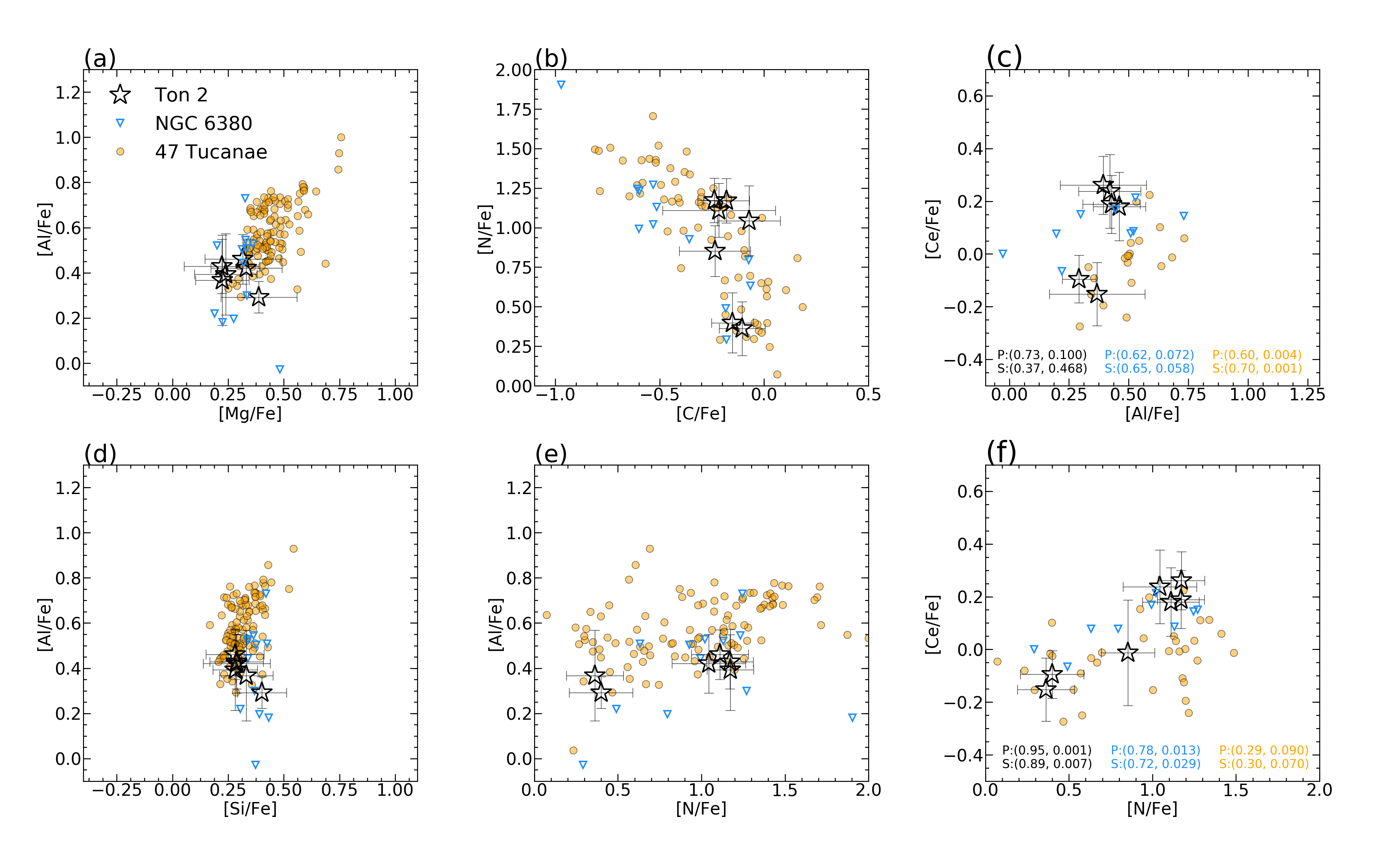}
		\caption{{\bf Combined light-, odd-Z, $\alpha$-, and \textit{s}-process elements.} Panel (a)--(f): [Mg/Fe]--[Al/Fe], [C/Fe]--[N/Fe], [Al/Fe]--[Ce/Fe], [Si/Fe]--[Al/Fe], [N/Fe]--[Al/Fe], and [N/Fe]--[Ce/Fe] distributions for Ton~2 (black asterisks), 47~Tucanae (orange circles) from \citet{Meszaros2020}, and NGC~6380 (blue inverted triangles) from \citet{Fernandez-Trincado2021d}. The typical uncertainties for Ton~2 members are also shown. In panels (c) and (f), Pearson's (P: first row of the annotation) and Spearman's (S: second row of the annotation) coefficients (first entry) and p$-$values (second entry) are indicated for Ton~2 (black), NGC~6380 (blue), and 47~Tucanae (orange). As in Fig. \ref{Figure4}, the abundance ratios shown here have been determined by adopting photometric atmospheric parameters. 
		}
		\label{Figure51}
	\end{center}
\end{figure*}

\section{Elemental abundances}
\label{elements}

The APOGEE-2 spectra provide access to 26 chemical species. However, most of the atomic and molecular lines are very weak and heavily blended, in some cases too much to produce reliable abundances in cool, relatively metal-rich bulge GC stars. For this reason, and after a careful visual inspection of all our spectra, we provide reliable abundance determinations for eleven selected chemical species, belonging to the iron-peak (Fe, Ni), odd-Z (Al), light- (C, N), $\alpha$-elements (O, Mg, Si, Ca and Ti), and \textit{s}-process (Ce) elements. 

It is important to note that we did not include sodium in our analysis, which is a typical species to separate GC populations, as this relies on two atomic lines (Na I: 1.6373$\mu$m and 1.6388$\mu$m) in the \textit{H}-band of the APOGEE-2 spectra.  These lines are generally very weak and heavily blended by telluric features, and thus not able to produce reliable [Na/Fe] abundance determinations in GCs with the typical T$_{\rm eff}$ and metallicities as Ton~2. For this reason, we place greater emphasis on the elemental abundances of Al, Mg, C, N, and O, which are typical chemical signatures to distinguish (at least) two main groups of  stars with different chemical composition in the multiple population phenomenon \citep[see, e.g.,][and references therein]{Ventura2016, Pancino2017, Tang2017, Tang2018, 6522_2019, Masseron2019, Meszaros2020, UKS1_2020, 6723_2021, M54_2021, VVVCL001_2021, Geisler2021, Meszaros2021, Romero-Colmenares2021, Tang2021}.

Overall, Figure \ref{Figure4} reveals that almost all of the chemical species examined so far in Ton~2 stars exhibit a very similar chemical enrichment as that of 47~Tucanae (disk GC) from \citet{Meszaros2020} and NGC~6380 (bulge GC) from \citet{Fernandez-Trincado2021d}.

\subsection{The iron-peak elements: Fe and Ni}

We measure a mean metallicity of $\langle$[Fe/H]$\rangle = -0.70\pm 0.05$($1\sigma$) $\pm$ 0.03(${\rm std}/\sqrt{\rm N}$), with a dispersion of  0.07 $\pm$ 0.02(${\rm std}/\sqrt{\rm 2\times N}$) dex.

Table \ref{Table2} lists a large total [Fe/H] range (0.22 dex) for Ton~2, which is mainly produced by the high metallicity ($-0.55$) of the star 2M17360837$-$3833312 (hereafter cluster outlier), whose deviation from the mean is on the order of the typical internal errors ($\sigma_{\rm [Fe I/H]^{*}} \sim 0.11$ -- $0.21$). Finally, the [Fe/H] abundance ratios listed in Table \ref{Table2} show that the observed dispersion agrees well with the measurement errors, so we find no evidence for a statistically significant metallicity spread. Our [Fe/H] is slightly more metal poor than the [Fe/H]$\approx -0.6$ estimated by \citet{Cote1999}, and significantly more metal poor than the tabulated by \cite{Vasquez2018} from the CaT reduced equivalent width, [Fe/H]$_{\rm S12} = -0.26$. Our [Fe/H] is also slightly more metal rich than the value tabulated by \citet{Dias2016b}, [Fe/H]$= -0.73\pm0.13$, which originates from the CMD-based [Fe/H] determination of \citet{Bica1996}, and is in reasonable agreement with the value reported in \citep{Harris1996}, [Fe/H]$=-0.7$. As our [Fe/H] determination was estimated directly from Fe I atomic lines and high-resolution spectra, this is likely more precise than the literature estimations. 
 
The relatively high metallicity we derive makes Ton~2 an interesting object, as GCs at comparable metallicity have been poorly studied within a $\sim$4 kpc Galactocentric radius. There are a handful of them with a lack of detailed chemical information (Terzan~2, Terzan~3, NGC~6637)--\citep{Harris1996, Baumgardt2018}, with the exception of a few chemical species examined in Terzan~2 \citep[][]{Geisler2021}. For this reason, in Figure \ref{Figure4} we compare our chemical makeup of Ton~2 to that of 47~Tucanae, taken from \citet{Meszaros2020} and NGC~6380 recently examined with APOGEE-2S data by \citet{Fernandez-Trincado2021d}, both clusters with a similar metallicity as Ton~2. Thus, Ton~2 complements this lack of information within a few kpc from the Galactic center. 

Regarding the other iron-peak elements we examined, nickel (Ni) is on average slightly super-solar ($\langle$[Ni/Fe]$\rangle= +0.07\pm0.05$) with a very small dispersion, $\sigma_{\rm [Ni/Fe]}<$0.07 dex, and a relatively high star-to-star spread ($\sim$0.23 dex), which is mainly produced by the cluster outlier. Therefore, beyond this cluster outlier, within uncertainties, we do not detect a significant spread in this element. It is also probable that the cluster outlier could be a variable star, which could explain the high offset in [Fe/H] and [Ni/Fe] compared to other cluster members, as there is some evidence that variability affects the measurement of the iron abundance in some way \citep[see,][ for instance]{Munoz2018}, causing an offset with respect to the cluster mean. Thus detailed photometric and spectroscopic analysis of this star is needed to investigate for possible variability effects on the metallicity and nickel derivation. 

The average $\langle$[Ni/Fe]$\rangle$ abundance ratio in Ton~2 is slightly higher than that observed in extragalactic environments at similar metallicity as Ton~2 \citep[see Fig. 4 in][]{Romero-Colmenares2021}, but a feature common to other bulge GCs at similar metallicity such as NGC~6380 (see Figure \ref{Figure4}), thus supporting a genuine Galactic origin for Ton~2. This is also well-supported by its dynamical history (see Section \ref{model}).

\subsection{The odd-Z element: Al}

We find that Ton~2 exhibits a mean aluminium enrichment of $\langle$[Al/Fe]$\rangle = +0.39\pm 0.04$, with a dispersion of 0.05$\pm$0.01 dex. We did not find a strong variation in [Al/Fe] beyond the typical errors. As before, the cluster outlier is the star with the lowest [Al/Fe] abundance ratio in our sample. Unfortunately, beyond this cluster outlier, and due to the small sample size of Ton~2, we are unable to identify any clear [Al/Fe] variation in Ton~2. It is worth to mentioning that both the Pearson's and Spearman's rank coefficients presented in Figure \ref{Figure51} do not support the prevalence of  an apparent Al-Ce correlation in Ton~2. Thus, with the present data, we do not find evidence for a clear [Al/Fe]--[Ce/Fe] correlation in Ton~2.

Moreover, it is important to notice that the possible absence of a spread in Al appears consistent with expectations of AGB nucleosynthesis \citep{Ventura2008, Ventura2016, Karakas2014, Crestani2019}, and its expected low production in relatively high metallicity GC stars such as observed in Ton~2. Figure \ref{Figure51}(a) shows no clear signs of an Mg-Al anti-correlation in Ton~2, which is consistent with no net production of these elements, rather, just the result of the conversion of Mg into Al during the Mg-Al cycle \citep[e.g.,][]{Pancino2017}.  

Figure \ref{Figure51}(d) also reveals that Ton~2 exhibits a mean $\langle$[Si/Fe]$\rangle = +0.33 \pm0.06$, comparable to the aluminium enrichment as expected from the breakout in the Mg--Al cycle \citep{Pancino2017}, i.e., with Al-rich stars that also present enrichment in Si. However, we find a small [Si/Fe] scatter ($<0.13$ dex) for Ton~2, with a non statistically significant Si-Al correlation, as the [Al/Fe] and [Si/Fe] abundance ratios of stars in Ton~2 concentrate tightly around the mean, a typical behavior observed in almost all the relatively high-metallicity ([Fe/H]$> -1$) Galactic GCs \citep[see, e.g.,][]{Pancino2017, Masseron2019, Meszaros2020, Geisler2021}.

\subsection{The light-elements: C and N}

Ton~2 exhibits a high enrichment in nitrogen, with a mean $\langle$[N/Fe]$\rangle = +0.87\pm0.39$, and a large star-to-star spread of $+0.81$ dex. Almost all the stars examined in Ton~2 are enriched in nitrogen with no significant [C/Fe] spread, ($\lesssim +0.17$ dex). Figure \ref{Figure51}(b) and (e) do not reveal a statistically significant C-N anticorrelation and N-Al correlation, however the chemical trends of these elements are similar to those observed in 47~Tucanae and NGC~6380, but with (at least) two groups of stars, likely compatible with a first stellar generation ([N/Fe]$\lesssim +0.5$ --including the cluster outlier, and the second stellar generations ([N/Fe]$\gtrsim+0.5$). This study reveals that a significant fraction of the stars with enhanced [N/Fe] abundances well above Galactic levels ([N/Fe]$\gtrsim+0.5$) populate Ton~2, a feature that is typical of stars in bulge clusters such as NGC~6380 \citep{Fernandez-Trincado2021d}, and a clear indication of multiple stellar populations \citep[see, e.g.,][]{Schiavon2017, UKS1_2020, 6723_2021, VVVCL001_2021, Geisler2021}. 

\subsection{The $\alpha$-elements: O, Mg, Si, Ca and Ti}

Figure \ref{Figure4} and Table \ref{Table2} show that Ton~2 exhibits a considerable $\alpha$-element enhancement, with mean values ranging from $+$0.27 to $+$0.39 ([O/Fe], [Mg/Fe], [Si/Fe], [Ca/Fe] and [Ti/Fe]), and in reasonable agreement with GCs of similar metallicity such as 47~Tucanae and NGC~6380, with small star-to-star spread (see Table \ref{Table2}), within our typical uncertainties, with the exception of the cluster outlier. Oxygen is the only $\alpha$-element that is slightly higher than other $\alpha$-element species, which exhibits homogeneity, at least as far as the observed dispersion is concerned. 

 The $\alpha$-elements in Ton~2 are overabundant compared to the Sun, which is  a feature common to almost all Galactic GCs and field stars at similar metallicity as the Ton~2 stars. So, according to its $\alpha$-element content, Ton~2 is similar to other Galactic GCs at similar metallicity, such as 47~Tucanae and NGC~6380.

\begin{figure}
\begin{center}
\includegraphics[width=90mm]{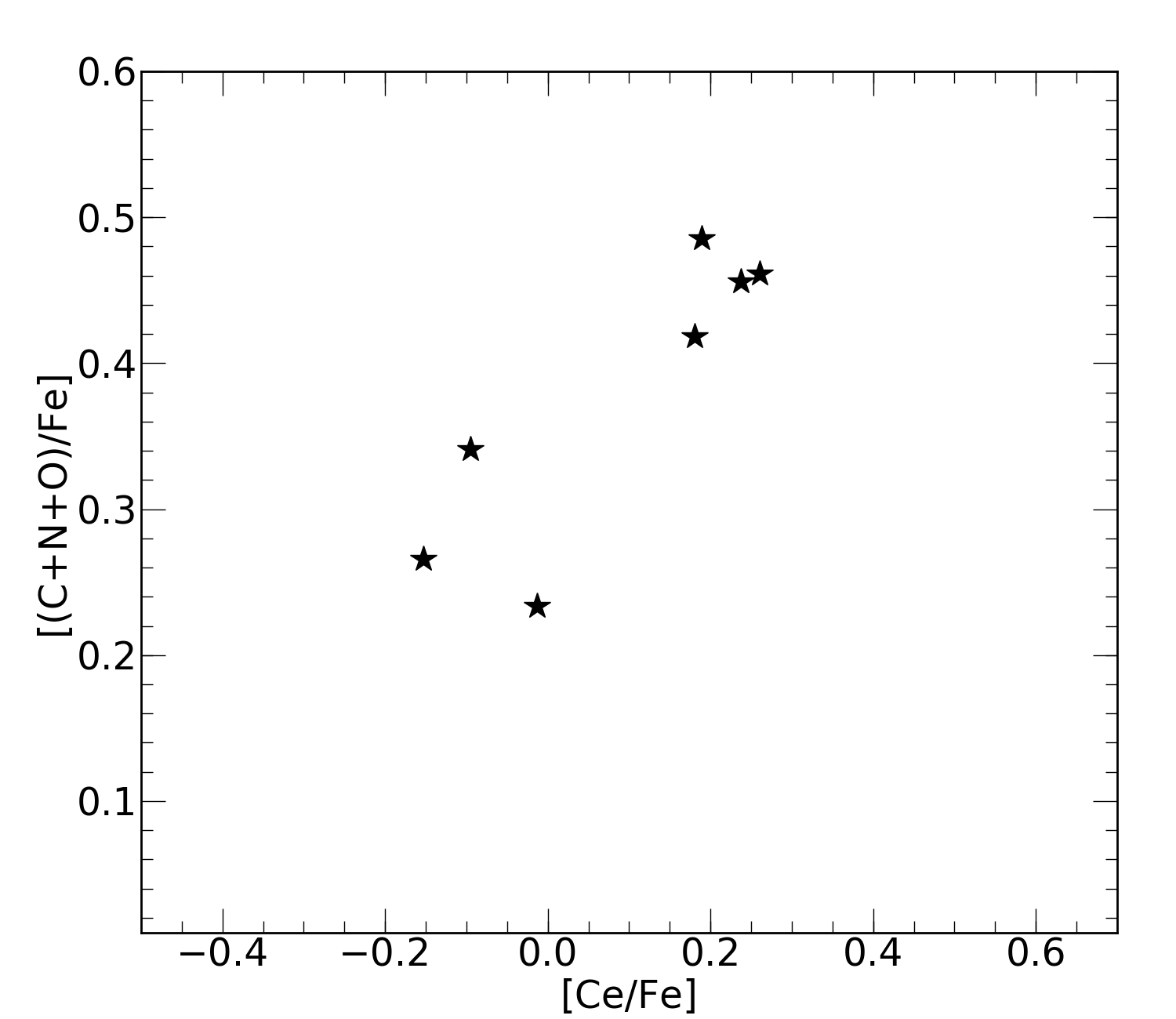}
\caption{[(C$+$N$+$O)/Fe] as a function of [Ce/Fe] for Ton~2 stars.}
\label{Figure9}
\end{center}
\end{figure}

\subsection{The \textit{s}-process element: Ce}

We find a mean $\langle$[Ce/Fe]$\rangle = +0.09 \pm 0.17$, which is slightly overabundant compared to the Sun, but comparable to the Ce levels observed in other Galactic GCs at similar metallicity \citep[see, e.g.][]{Masseron2019, Meszaros2020}. The \textit{s}-process element Ce has an observed spread ($\gtrsim+0.41$ dex) that exceeds the observational uncertainties. More importantly, Figure \ref{Figure51}(c) and (f ) reveal the presence of Ce abundances which are remarkably correlated with N, as indicated by the Pearson and Spearman correlation test in the same figure. The low p$-$value indicate that the observed correlation is unlikely due to random chance. 

This  atypical feature, has been observed only in the bulge GC NGC~6380 \citep{Fernandez-Trincado2021d}, with the exception of 47~Tucanae, which exhibits only a marginal correlation of Ce with N but a very strong correlation of Ce with Al. Thus, we believe that the \textit{s}-process enrichment in Ton~2 has been produced by different progenitors, possibly by low-mass AGB stars \citep[see, e.g.,][]{Ventura2009}. In the high-metallicity regime of Ton~2, the clear increase of the Ce abundance as N  increases supports this assertion. Ton~2 is the second case of a relatively high-metallicity bulge GC where a clear N-Ce correlation has been detected, confirming that bulge GCs at this metallicity regime have likely experienced a different chemical evolution with respect to the bulk of MW GCs. However, a consensus interpretation of the origin for such N-Ce or Al-Ce correlations is still lacking.

In Figure \ref{Figure9} the [(C$+$N$+$O)/Fe] abundance is represented as a function of [Ce/Fe]. The Ton~2 population split is evident; the Ce-rich stars have on average a higher  [(C$+$N$+$O)/Fe] abundance, with $\Delta^{\rm rich}_{\rm poor}$[(C$+$N$+$O)/Fe]$\sim +0.15$. This correlation between Ce and the CNO abundance sum strengthens the idea that relatively low mass AGB stars \citep[$\sim$3 M$_{\odot}$,][]{Ventura2009} have likely polluted the intra-cluster medium with \textit{s}-process elements, similar as in other GCs such as NGC~1851 \citep{Yong2009}, M~22 \citep{Marino2011}, and $\omega$ Cen \citep{Marino2012}. This finding indicates the prevalence of the correlation of Ce with CNO in GCs as relatively metal rich as Ton~2.

\begin{figure}
\begin{center}
\includegraphics[width=90mm]{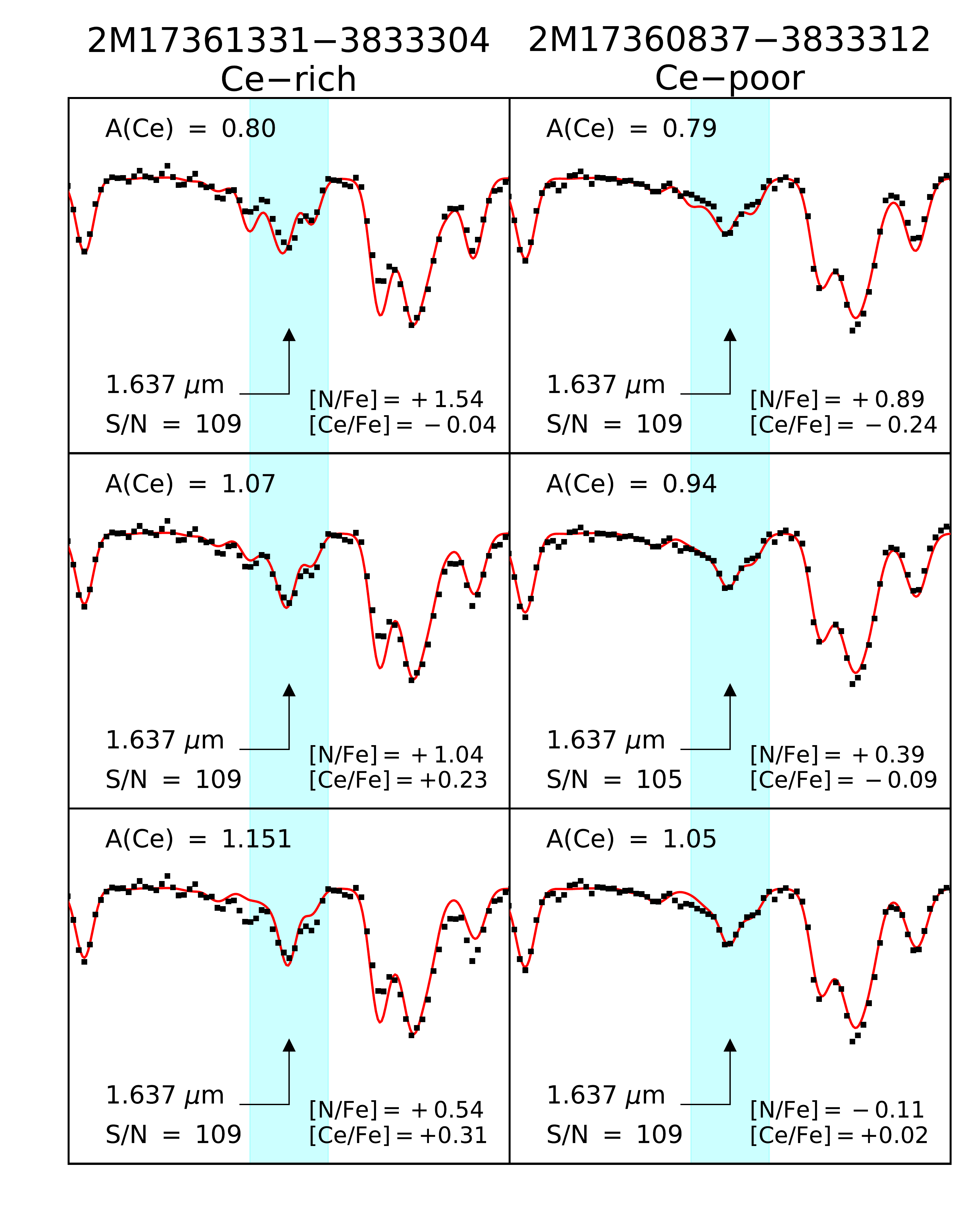}
\caption{Comparison of synthetic spectra (red lines) to the observed spectra (black symbols) for a Ce-rich and Ce-poor member of Ton~2. In each row we show the effect of changing the abundance of [N/Fe]. The second row shows the case of the best-determined [N/Fe] and [Ce/Fe] abundaces ratios listed in Table \ref{Table2}.}
\label{Figure12}
\end{center}
\end{figure}

Figure \ref{Figure12} shows a portion of the spectra around the Ce II line at 1.637 $\mu$m (cyan bands) of two Ton~2 members: the Ce-rich star 2M17361331$-$3833304, and the Ce-poor star 2M17360837$-$3833312, with similar atmospheric parameters (see Table \ref{Table2}), and the success of the \texttt{BACCHUS} fitting procedure. The first and third row in this figure manifests the effect on the synthetic spectrum around the Ce II line in changing the [N/Fe] abundance ratio listed in Table \ref{Table2} by $\pm0.5$ dex. Thus, by lowering or increasing the [N/Fe]  abundance ratio by $\pm0.5$ dex the Ce II line is not well-reproduced by the fit, however, the second row shows the success of the determined [N/Fe] and [Ce/Fe] on reproducing the profile of the observed Ce II line.

\begin{table*}
	\begin{small}
		\begin{center}
			\setlength{\tabcolsep}{1.mm}  
			\caption{\texttt{BACCHUS} elemental abundances for the observed stars.}
			\begin{tabular}{lccccccccccccccc}
				\hline
				\hline
				APOGEE-ID & S/N &   T$_{\rm eff}$ & $\log$ \textit{g}  &   $\xi_{t}$   & [C/Fe] & [N/Fe] & [O/Fe] & [Mg/Fe]  & [Al/Fe] & [Si/Fe] & [Ca/Fe] & [Ti/Fe]  & [Fe/H]  & [Ni/Fe]  & [Ce/Fe]\\
				& pixel$^{-1}$ &   K & cgs &   km s$^{-1}$   &  &  &  &   & &  & & &   & &  \\
				\hline
				\hline
				2M17360034$-$3835151 &  275 & 3530 &  0.12 &  2.64   & $-$0.23 & $+$0.85  & $+$0.20  & $+$0.30   & ...                & $+$0.41   & $+$0.22   &  $+$0.17   & $-$0.75 &    $+$0.09 & $-$0.01   \\
				& & & & & (0.17) &  (0.16) &  (0.19) & (0.23) &  ... &  (0.15) &  (0.08) &  (0.13) & (0.17) &  (0.15) &  (0.20) \\
				2M17361421$-$3834371 &  224 & 3730 &  0.48 &  1.95   & $-$0.10 & $+$0.36  & $+$0.37  & $+$0.22   & $+$0.36   & $+$0.33   & $+$0.17   &  $+$0.21   & $-$0.65 &    $+$0.01 & $-$0.15   \\
				& & & & & (0.11) &  (0.17) &  (0.23) & (0.12) &  (0.20) &  (0.12) &  (0.14) &  (0.16) & (0.21) &  (0.09) &  (0.12) \\
				2M17361150$-$3832114 &  123 & 4100 &  1.15 &  2.53   & $-$0.23 & $+$1.17  & $+$0.40  & $+$0.23   & $+$0.39   & $+$0.28   & $+$0.32   &  $+$0.39   & $-$0.73 &    $+$0.09 &    $+$0.26   \\
				& & & & & (0.08) &  (0.14) &  (0.15) & (0.14) &  (0.18) &  (0.10) &  (0.08) &  (0.19) & (0.15) &  (0.12) &  (0.11) \\
				2M17360837$-$3833312 &  109 & 4139 &  1.22 &  1.59   & $-$0.15 & $+$0.39  & $+$0.46  & $+$0.38   & $+$0.29   & $+$0.40   & $+$0.17   &  $+$0.22   & $-$0.55 &     $-$0.07 & $-$0.09   \\
				& & & & & (0.10) &  (0.19) &  (0.16) & (0.17) &  (0.07) &  (0.11) &  (0.13) &  (0.16) & (0.11) &  (0.14) &  (0.09) \\
				2M17361331$-$3833304 &  105 & 4139 &  1.22 &  2.27   & $-$0.07 & $+$1.04  & $+$0.45  & $+$0.33   & $+$0.42   & $+$0.28   & $+$0.27   &  $+$0.30   & $-$0.74 &    $+$0.10 &    $+$0.23   \\
				& & & & & (0.15) &  (0.22) &  (0.17) & (0.16) &  (0.13) &  (0.15) &  (0.10) &  (0.20) & (0.19) &  (0.11) &  (0.14) \\
				2M17355890$-$3834199 &   77 & 4224 &  1.37 &  2.38   & $-$0.18 & $+$1.16  & $+$0.44  & $+$0.22   & $+$0.42   & $+$0.28   & $+$0.36   &  $+$0.38   & $-$0.77 &    $+$0.15 &    $+$0.18   \\
				& & & & & (0.11) &  (0.14) &  (0.16) & (0.17) &  (0.12) &  (0.12) &  (0.08) &  (0.22) & (0.14) &  (0.15) &  (0.11) \\
				2M17360681$-$3834336 &   63 & 4352 &  1.61 &  1.93   & $-$0.21 & $+$1.10  & $+$0.37  & $+$0.31   & $+$0.46   & $+$0.28   & $+$0.34   &  $+$0.33   & $-$0.69 &    $+$0.09 &    $+$0.18   \\	
				& & & & & (0.27) &  (0.17) &  (0.17) & (0.17) &  (0.11) &  (0.13) &  (0.05) &  (0.19) & (0.13) &  (0.13) &  (0.13) \\			
				\hline
				\hline
				\rowcolor{pink}
				\textcolor{black}{\bf Mean}   & ...  & ...  & ...  & ...   & \textcolor{black}{ $\bf  -0.17$}    & \textcolor{black}{\bf $+$0.87}  & \textcolor{black}{\bf $+$0.39} &  \textcolor{black}{\bf $+$0.29}  & \textcolor{black}{\bf $+$0.39}  & \textcolor{black}{\bf $+$0.33} & \textcolor{black}{\bf $+$0.27}   &\textcolor{black}{\bf $+$0.29} & \textcolor{black}{$\bf  -0.70$} & \textcolor{black}{\bf $+$0.07}  & \textcolor{black}{\bf $+$0.09} \\ 
				$1\sigma$        &  ... &  ... &  ... &  ...  & $0.07$   & $0.39$ &  $0.05$ & $0.06$ & $0.04$ &  $0.06$ & $0.09$ &  $0.09$ & $0.05$ & $0.05$& $0.17$   \\   
				std          &  ... &  ... &  ... &  ...  & $0.06$   & $0.33$ &  $0.08$ & $0.06$ & $0.05$ &  $0.05$ & $0.07$ &  $0.08$ & $0.07$ & $0.07$& $0.16$   \\   
				spread   &  ... &  ... &  ... &  ...  & $0.17$   & $0.81$ &  $0.26$ & $0.17$ & $0.17$ &  $0.13$ & $0.19$ &  $0.22$ & $0.22$ & $0.23$& $0.41$   \\   		
				\hline
				\hline
			\end{tabular}  \label{Table2}
		\end{center}
			\raggedright{{\bf Note:} The mean elemental abundances, and 1$\sigma$ error, the standard deviation (std), and spread are shown for the full sample. All of the listed elemental abundances have been scaled to the Solar reference value from \citet{Asplund2005}. $1\sigma $ is defined as (84$^{\rm th}$ percentile  $-$ 16$^{\rm th}$ percentile)/2. The estimated uncertainties ($\sigma_{total}$) indicated inside parentheses were computed in the same manner as described in \citet{6522_2019}.}
	\end{small}
\end{table*}   

\section{Rough estimate of the cluster mass}
\label{mass}

The precise APOGEE-2 RV information for our twelve stars were combined with other existing RV measurements of  Ton~2 members from \citet{Baumgardt2019}. This yields a unique collection of 22 likely members of Ton~2 with RV information. For our twelve APOGEE-2 sources, we find a mean radial velocity of $-178.61\pm0.86$ km s$^{-1}$ with a velocity dispersion of  2.99$\pm$0.61 km s$^{-1}$, while for other remaining stars with RV from the literature we find a mean radial velocity of $-183.08\pm1.74$ with a velocity dispersion of $5.51\pm1.23$ km s$^{-1}$, in good agreement, but with an apparent systematic difference of $\sim 4.47$ km s$^{-1}$ between both samples, comparable to the internal dispersion. 

\begin{figure}
	\begin{center}
		\includegraphics[width=90mm]{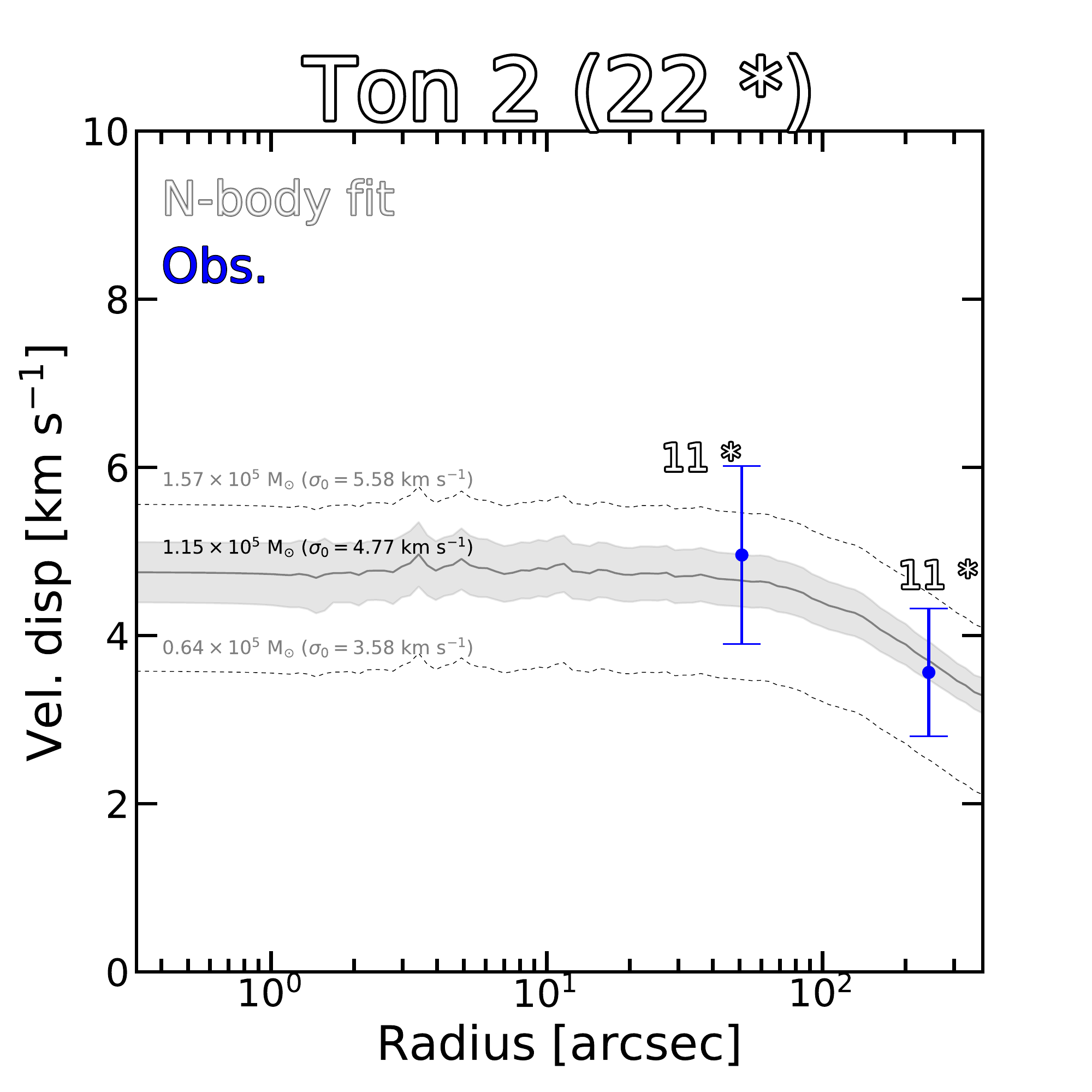}
		\caption{{\bf Velocity dispersion profile of Ton~2.} Line-of-sight velocity dispersion versus radius for our target cluster stars from APOGEE-2 plus \citet{Baumgardt2019} data set. The error bars are refered as $\sigma_{\rm RV}/\sqrt{2\times {\rm N}}$, with N the number of stars per bin as indicated in the plot. The prediction of the best-fitting updated (private communication) \textit{N}-body model from \citep{Baumgardt2018} and \citet{Baumgardt2019} is shown as a solid-grey line, and the light-grey shaded region indicates the 1$\sigma$ uncertainty from the fit. The upper and lower mass limit according to the observed error bars is indicated by the black dashed lines.}
		\label{Figure5}
	\end{center}
\end{figure}

With this relatively large sample of stars with RV information, and by appropriately taking account possible systematics of the combined sample, we match the line-of-sight dispersion profiles to the updated version of \textit{N}-body simulations (private communication, H. Baumgardt) of  Ton~2 from \citet{Baumgardt2018, Baumgardt2019}, as shown in Figure \ref{Figure5}, and thus determine the most likely mass of the cluster from kinematic constraints. 

We adopted two radial bins (with bin centers of 0.75\arcmin and 3.75\arcmin), chosen to ensure that at least eleven stars were in each bin, resulting in the two points shown in Figure \ref{Figure5}. We find $\sigma_{0}\sim4.77^{+5.13}_{-4.40}$ km s$^{-1}$, yielding a present-day estimated mass of $\sim1.15^{+1.33}_{-0.98}\times$10$^{5}$ M$_{\odot}$ within a projected half-light radius of r$_{\rm h,l}$ $= 2.89$ pc taken from \citep{Baumgardt2019}. However, it is important to note that the observed large error bars shown in Figure \ref{Figure5} yields a mass uncertainty in the range between 0.64 -- 1.57  $\times 10^{5}$ M$_{\odot}$, which could be better constrained with further observations.

\section{Ton~2 orbit}
\label{model}

Orbits for Ton~2 have been predicted with the Galactic dynamic model \texttt{GravPot16}\footnote{https://gravpot.utinam.cnrs.fr}. The Galactic configuration is the same as described in \citet{Fernandez-Trincado2020}, except for the bar pattern speeds, for which we adopted the recommended value of $\Omega_{\rm bar} = 41$ km s$^{-1}$ kpc$^{-1}$ \citep[see, e.g.,][]{Sanders2019}, with an uncertainty of 10 km s$^{-1}$ kpc$^{-1}$ .

\begin{figure*}
	\begin{center}
		\includegraphics[width=95mm]{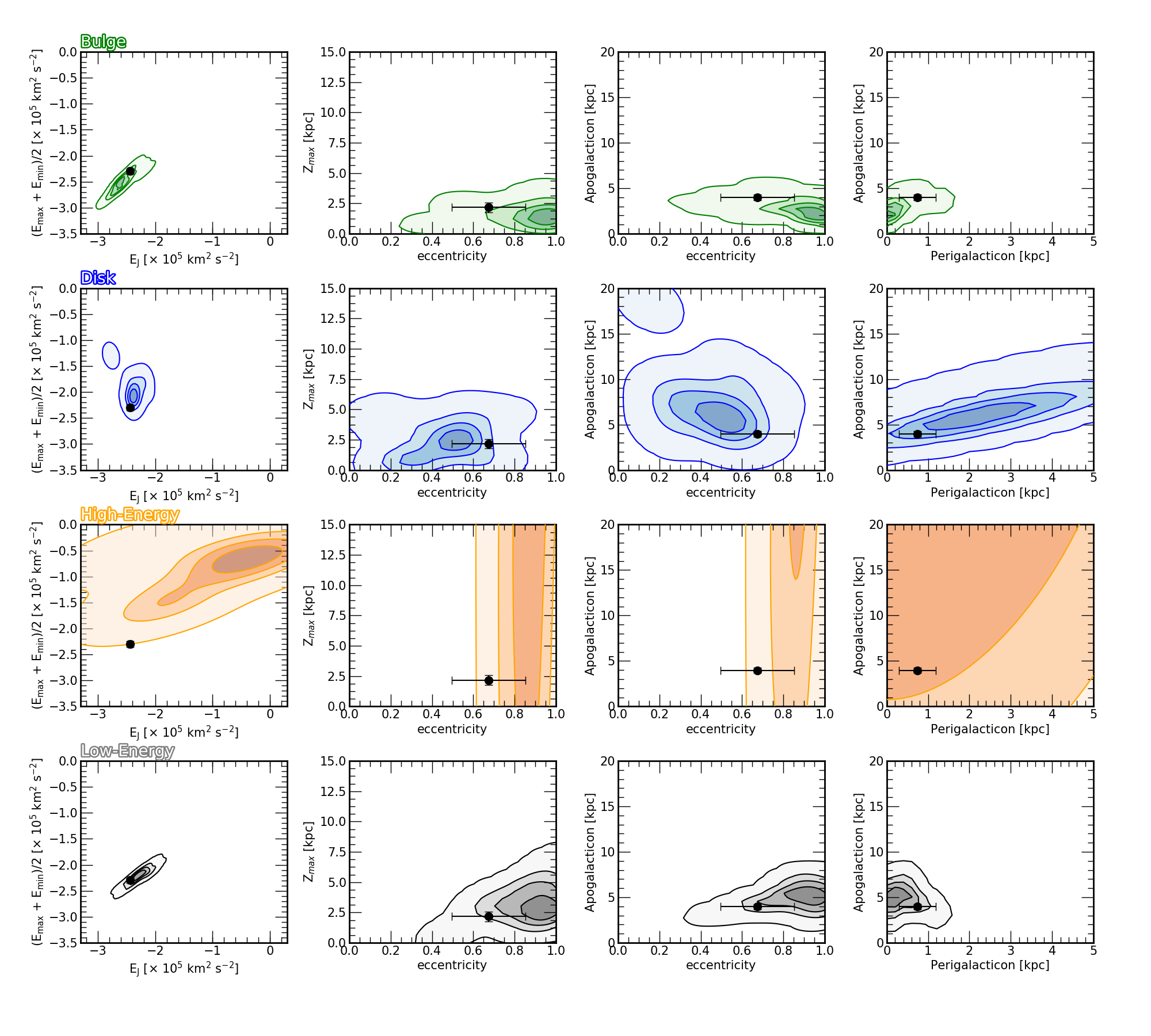}\includegraphics[width=85mm]{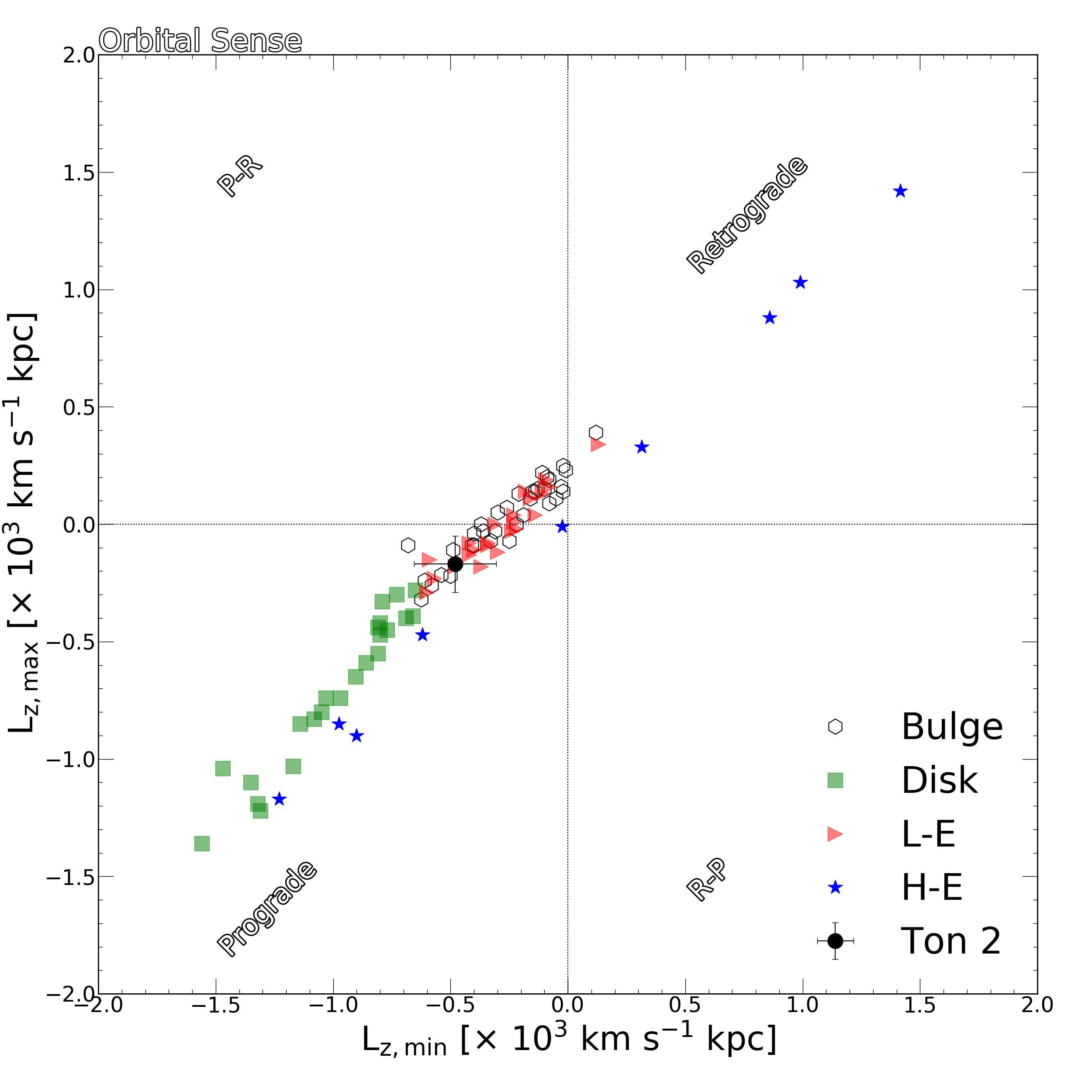}
		\caption{{\bf Dynamical properties of Ton~2.} Left panel: Kernel Density Estimation (KDE) models of the characteristic orbital energy ((E$_{\rm max}$ + E$_{\rm min}$)/2), the orbital Jacobi energy (E$_{\rm J}$), orbital pericenter and apocenter, orbital eccentricity, and maximum vertical height above the Galactic plane determined from our Galactic model (see Section \ref{model}) for GCs with a Galactic origin according to the classification from \citet{Massari2019}. Ton~2 is highlighted with black dot symbols. Right panel: The minimum and maximum value of the z-component of the angular momentum in the inertial frame, indicating the regions dominated by prograde and retrograde orbits, and those that have prograde and retrograde (P$-$R) or retrograde and prograde (R$-$P) orbits at the same time.}
		\label{Figure6}
	\end{center}
\end{figure*}

We have employed a simple Monte Carlo approach and the Runge-Kutta algorithm of seventh to eighth order, as elaborated by \citet{fehlberg68}. As input parameters we adopted the following observables: (\textit{i}) RV $= -178.6$ km s$^{-1}$ (see Section \ref{mass}) with a dispersion of 2.99 km s$^{-1}$; (\textit{ii}) absolute proper motions $\mu_{\alpha}\cos(\delta)=-5.913\pm0.031$ mas yr$^{-1}$ and $\mu_{\delta}=-0.758\pm0.028$ mas yr$^{-1}$ from \citet{Vasiliev2021}; and (\textit{iii}) a heliocentric distance of $d_{\odot} = 7.76$ kpc (see subsection \ref{isochrone}) with an assumed uncertainty on the order of 10\% . The uncertainties in the input data (e.g. $\alpha$, $\delta$, distance, proper motions, and RV errors) were randomly propagated as 1$\sigma$ variations in a Gaussian Monte Carlo resampling. Thus, we ran orbits computed backwards in time over 3 Gyr. The median value of the orbital elements were found for ten thousand realizations, with uncertainty ranges given by the 16$^{\rm th}$ and 84$^{\rm th}$ percentile. The resulting orbital elements are portrayed in Figure \ref{Figure6}. Figure \ref{Figure6} (right panel) shows the minimal and maximum value of the z-component of the angular momentum in the inertial frame \{L$_{\rm z, min}$, L$_{\rm z, max}$\}, since this quantity is not conserved in a model with non-axisymmetric structures like \texttt{GravPot16}. In the case of Ton~2, L$_{\rm z, min}$ and L$_{\rm z, max}$ are close to each other within the uncertainties, thus confirming the genuine prograde nature of  Ton~2, as shown on the right panel in Figure \ref{Figure6}.

Overall, our orbital analysis suggests that Ton~2 is confined in a prograde, high-eccentricity ($\sim 0.67 \pm 0.18$) and radial (r$_{\rm peri} \lesssim 0.74 \pm 0.45 $ kpc) orbit, with relatively low vertical excursions  from the Galactic plane, Z$_{\rm max} \lesssim  2.16 \pm 0.40 $ kpc, circulating (r$_{\rm apo}\lesssim 3.97 \pm 0.18$ kpc) within the co-rotation radius. This orbital configuration suggests that Ton~2 is not uncommon among other GCs in the inner Galaxy. Figure \ref{Figure6} (left panel) shows that Ton~2 shares similar orbital properties with those GCs in the bulge, disk, and the class of GCs with low binding energy (L-E) \citep[see, e.g.,][]{Massari2019}. However, the right panel of Figure \ref{Figure6} shows that Ton~2 is on the boundary between the bulge and L-E. 

Further inspection on the classification proposed in \citet[][]{Massari2019}, in combination with the relatively high metallicity of Ton~2 make this object a potential candidate formed in situ, and part of the Main-Progenitor group (as L-E GCs at metallicities like Ton~2 are less common). It is likely a bulge GC, such as suggested by its  angular momentum distribution, as can be appreciated in Figure \ref{Figure6}--\{L$_{\rm z, min}$, L$_{\rm z, max}$\}, rather than a disk GC as suggested by \citet{Perez-Villegas2020}. We also examined the possible link of Ton~2 with accreted GCs in the same manner as presented in Fig. 7 in \citet[][]{Romero-Colmenares2021}. However, we find that Ton~2 does not share orbital properties that are linked with any of the GCs group with an accreted origin \citep{Massari2019}, therefore we discard this possibility for Ton~2. 

It is important to mention that one reason for the discrepant classification between \citet[][]{Perez-Villegas2020} ($d_{\odot} = 6.40 \pm 0.64$ kpc) and our work is due to the adoption of a different heliocentric distance. A distance close to the Solar position maximizes the probability of  a disk classification, however, the good agreement between our estimated distance and the recent revised estimation by \citet{Baumgardt2021} favors a larger heliocentric distance for Ton~2 (see Section \ref{isochrone}).

The orbit configuration of Ton~2 places it among the potential clusters that have likely experienced important mass loss \citep[e.g.,][]{Leon2000, Moreno2014, Minniti2018, Kundu2019, Kundu2021} from early epochs, as a result of the strong tidal field in the inner galaxy, making Ton~2 a potential progenitor candidate for the chemically unusual stars identified in the inner Galactic field at similar metallicity as Ton~2 \citep[see, e.g.,][]{Fernandez-Trincado2016, Fernandez-Trincado2017, Binary_2019, Fernandez-Trincado2019b, Fernandez-Trincado2019, Fernandez-Trincado2020c, Fernnadez-Trincado2020_Aluminum, Fernandez-Trincado2020, Recio-Blanco2017}. 

\section{Conclusions}
\label{conclusions}

We present the first detailed elemental-abundance analysis for seven stars belonging to the heavily reddened globular cluster Ton~2 in the direction of the inner Galaxy. We examined eleven chemical species belonging to the  light- (C and N), $\alpha$- (O, Mg, Si, Ca, and Ti), iron-peak (Fe and Ni), odd-Z (Al), and \textit{s}-process (Ce) elements. Overall, the chemical species examined so far in Ton~2 are in agreement with other Galactic GCs at similar metallicity \citep[e.g.,][]{Meszaros2020}. The main conclusions of this paper are the following:

\begin{itemize}
	
	\item Ton~2 exhibits a mean metallicity of $\langle$[Fe/H]$\rangle = -0.70\pm0.05$, with a star-to-star [Fe/H] spread that is comparable to the uncertainties, with one potential outlier.
	
	\item The Mg-Al anti-correlation is not seen in Ton~2, which is in line with other GCs at similar metallicity \citep[see, e.g.,][]{Pancino2017}. 
	
	\item We also find a significant variation in cerium, with a clear [Ce/Fe]--[N/Fe] and [Ce/Fe]--[(C$+$N$+$O)/Fe] correlation, and a significant spread in nitrogen ($>0.81$ dex), cerium ($>0.41$ dex), and modest spread  in the CNO sum, $\Delta^{\rm rich}_{\rm poor}$[(C$+$N$+$O)/Fe]$\sim +0.15$. We hypothesize that this feature could be produced by different progenitors, more likely by low-mass ($\sim$3 M$_{\odot}$) AGB stars \citep{Ventura2009}, similar to that detected in other \textit{s}-elements in GCs \citep[see, e.g.][]{Yong2009, Marino2011, Marino2012}. Ton~2 is the second known case in the relatively high-metallicity regime to exhibit this feature after NGC~6380 \citep{Fernandez-Trincado2021d}, suggesting that this phenomenon could be common among relatively high-metallicity bulge GCs, and Ce could be a good indicator for the prevalence of the multiple-population phenomenon.
	
	\item Ton~2 hosts a significant population of nitrogen-enriched stars indicating the prevalence of the multiple-population phenomenon in this cluster. Furthermore, the high nitrogen enrichment of Ton~2 makes this object a potential progenitor of the nitrogen-enhanced metal-rich field stars identified in the inner Galaxy \citep[e.g.,][]{Fernandez-Trincado2017, Schiavon2017b, Fernandez-Trincado2019, Fernandez-Trincado2019b, Fernandez-Trincado2020, Fernnadez-Trincado2020_Aluminum}. 
	
	\item We find that Ton~2 is likely as massive as $\sim1.15^{1.33}_{0.98}\times$10$^{5}$ M$_{\odot}$ and likely more massive in the past, having experienced significant mass loss from early epochs as a result of the strong tidal field in the inner Galaxy. Thus, Ton~2 could be one of the potential progenitors for the chemically unusual stars identified in the inner Galactic region at similar metallicity \citep[see, e.g.,][]{Fernandez-Trincado2017, Fernandez-Trincado2019, Fernandez-Trincado2019b, Fernandez-Trincado2020, Fernnadez-Trincado2020_Aluminum}.
	
	\item Ton~2 lies in a bulge-like (r$_{\rm apo}\lesssim 3.97$), prograde, high-eccentricity ($\sim 0.67$), and radial orbit (r$_{\rm peri} \lesssim 0.74 $ kpc), with relatively low vertical excursions  from the Galactic plane (Z$_{\rm max} \lesssim  2.16 $ kpc). We find that it is a high-metallicity cluster that {\it does not} live in the disk, as previously thought \citep{Perez-Villegas2020} but, rather, in the bulge region. We have caught Ton~2 near the pericentre of its orbit (R$_{\rm gal} \sim$1.29 kpc).
	
\end{itemize}

	\begin{acknowledgements}  
		The authors are grateful for the enlightening feedback from an anonymous referee. We warmly thank Holger Baumgardt for providing his more recent numerical \textit{N}-body modeling of the line-of-sight velocity dispersion of Ton~2.  D.G. gratefully acknowledges support from the Chilean Centro de Excelencia en Astrof\'isica y Tecnolog\'ias Afines (CATA) BASAL grant AFB-170002. D.G. also acknowledges financial support from the Direcci\'on de Investigaci\'on y Desarrollo de la Universidad de La Serena through the Programa de Incentivo a la Investigaci\'on de Acad\'emicos (PIA-DIDULS). D.G. and D.M. gratefully acknowledge support from the Chilean Centro de Excelencia en Astrof\'isica y Tecnolog\'ias Afines (CATA) BASAL grant AFB-170002. T.C.B. acknowledges partial support for this work from grant PHY 14-30152: Physics Frontier Center / JINA Center for the Evolution of the Elements (JINA-CEE), awarded by the US National Science Foundation. B.B. acknowledges grants from FAPESP, CNPq and CAPES - Financial code 001. E.R.G. acknowledges support from ANID PhD scholarship No. 21210330. C.M. thanks the support provided by FONDECYT Postdoctorado No.3210144.\\

        This work has made use of data from the European Space Agency (ESA) mission \textit{Gaia} (\url{http://www.cosmos.esa.int/gaia}), processed by the \textit{Gaia} Data Processing and Analysis Consortium (DPAC, \url{http://www.cosmos.esa.int/web/gaia/dpac/consortium}). Funding for the DPAC has been provided by national institutions, in particular the institutions participating in the \textit{Gaia} Multilateral Agreement.\\

        Funding for the Sloan Digital Sky Survey IV has been provided by the Alfred P. Sloan Foundation, the U.S. Department of Energy Office of Science, and the Participating Institutions. SDSS- IV acknowledges support and resources from the Center for High-Performance Computing at the University of Utah. The SDSS web site is www.sdss.org. SDSS-IV is managed by the Astrophysical Research Consortium for the Participating Institutions of the SDSS Collaboration including the Brazilian Participation Group, the Carnegie Institution for Science, Carnegie Mellon University, the Chilean Participation Group, the French Participation Group, Harvard-Smithsonian Center for Astrophysics, Instituto de Astrof\`{i}sica de Canarias, The Johns Hopkins University, Kavli Institute for the Physics and Mathematics of the Universe (IPMU) / University of Tokyo, Lawrence Berkeley National Laboratory, Leibniz Institut f\"{u}r Astrophysik Potsdam (AIP), Max-Planck-Institut f\"{u}r Astronomie (MPIA Heidelberg), Max-Planck-Institut f\"{u}r Astrophysik (MPA Garching), Max-Planck-Institut f\"{u}r Extraterrestrische Physik (MPE), National Astronomical Observatory of China, New Mexico State University, New York University, University of Notre Dame, Observat\'{o}rio Nacional / MCTI, The Ohio State University, Pennsylvania State University, Shanghai Astronomical Observatory, United Kingdom Participation Group, Universidad Nacional Aut\'{o}noma de M\'{e}xico, University of Arizona, University of Colorado Boulder, University of Oxford, University of Portsmouth, University of Utah, University of Virginia, University of Washington, University of Wisconsin, Vanderbilt University, and Yale University.\\
	\end{acknowledgements}


\end{document}